\documentclass[conference]{IEEEtran}
\IEEEoverridecommandlockouts
\usepackage{cite}
\usepackage{amsmath,amssymb,amsfonts}
\usepackage{algorithmic}
\usepackage{graphicx}
\usepackage{xurl}
\usepackage{fancyhdr} 
\usepackage{hyperref}  
\usepackage{textcomp}
\usepackage{xcolor}
\def\BibTeX{{\rm B\kern-.05em{\sc i\kern-.025em b}\kern-.08em
    T\kern-.1667em\lower.7ex\hbox{E}\kern-.125emX}}
\begin{document}

\title{Enhancing Meme Token Market Transparency: A Multi-Dimensional Entity-Linked Address Analysis for Liquidity Risk Evaluation\\
}

\author{\IEEEauthorblockN{1\textsuperscript{st} Qiangqiang Liu}
\IEEEauthorblockA{\textit{Risk Department} \\
\textit{Binance}\\
Dubai, United Arab Emirates \\
codi.l@binance.com}
\and
\IEEEauthorblockN{2\textsuperscript{st} Qian Huang}
\IEEEauthorblockA{\textit{Risk Department} \\
\textit{Binance}\\
Hong Kong, China \\
qian.huang@binance.com}
\and
\IEEEauthorblockN{3\textsuperscript{st} Frank Fan}
\IEEEauthorblockA{\textit{Risk Department} \\
\textit{Binance}\\
Hong Kong, China \\
frank.f@binance.com}
\and
\IEEEauthorblockN{4\textsuperscript{st} Haishan Wu}
\IEEEauthorblockA{\textit{AI Department} \\
\textit{Zand}\\
Dubai, United Arab Emirates \\
haishan.wu@zand.ae}
\and
\IEEEauthorblockN{5\textsuperscript{st} Xueyan Tang}
\IEEEauthorblockA{\textit{Suzhou Artificial Intelligence Research Institute} \\
\textit{Shanghai Jiao Tong University}\\
Suzhou, Jiangsu, China \\
mirror.tang@alumni.stanford.edu}
}

\maketitle

\begin{abstract}
Meme tokens represent a distinctive asset class within the cryptocurrency ecosystem, characterized by high community engagement, significant market volatility, and heightened vulnerability to market manipulation. This paper introduces an innovative approach to assessing liquidity risk in meme token markets using entity-linked address identification techniques. We propose a multi-dimensional method integrating fund flow analysis, behavioral similarity, and anomalous transaction detection to identify related addresses. We develop a comprehensive set of liquidity risk indicators tailored for meme tokens, covering token distribution, trading activity, and liquidity metrics. Empirical analysis of tokens like BabyBonk, NMT, and BonkFork validates our approach, revealing significant disparities between apparent and actual liquidity in meme token markets. The findings of this study provide significant empirical evidence for market participants and regulatory authorities, laying a theoretical foundation for building a more transparent and robust meme token ecosystem.
\end{abstract}


\section{Introduction}

In the context of rapid development in blockchain technology and cryptocurrencies, meme tokens have emerged as a
new class of digital assets, attracting widespread attention due to their unique characteristics and market dynamics. By 2024, meme coins have captured 11\% of the total cryptocurrency
market capitalization, exceeding \$120 billion USD, with tokens like WIF and TRUMP reaching multi-billion dollar valuations in remarkably short timeframes. This rapid capital inflow demonstrates the persistent strong market demand for meme tokens. Recent studies have highlighted the distinctive features of meme tokens within the cryptocurrency ecosystem\cite{b1,b2}: These tokens often originate from internet culture and social media trends, characterized by strong community engagement and viral propagation, as exemplified by TRUMP coin, which launched just before Trump's 2025 inauguration and surged from zero to a \$30 billion market cap within just 12 hours, driven primarily by political memes, the president's own social media posts. The rapid rise of the meme token market
reflects the innovative vitality of the cryptocurrency ecosystem, while also exposing the limitations of traditional financial analysis methods when faced with this new type of asset. Traditional financial analysis methods have shown significant limitations when applied to meme tokens\cite{b3}, highlighting the necessity for developing new risk assessment approaches.

These tokens typically exhibit the following characteristics: 1) High volatility: prices can fluctuate dramatically in short periods. 2) Social media-driven: value and liquidity are largely influenced by social media sentiment. 3) Small market capitalization and low liquidity: compared to mainstream cryptocurrencies, meme tokens usually have smaller market caps and limited liquidity. 4) Uncertain fundamentals: many meme tokens lack clear use cases or value propositions. 5) Potential market manipulation risk: due to small market size, they are susceptible to manipulation by large holders, as seen in LIBRA, where a president's endorsement caused a price surge followed by a crash costing investors \$250 million, and Broccoli, where concentrated holdings (60\% by top 10 addresses) enabled a pump-and-dump cutting prices in half within a day. These characteristics pose severe challenges to traditional financial analysis methods in evaluating meme tokens.

In the cryptocurrency ecosystem, Entity identification attributes multiple seemingly independent blockchain addresses to the same entity, whether an individual or organization. Our subsequent analysis primarily focusing on entity-linked addresses, a group of addresses whose keys are controlled by the same entity (individual, organization, or institution). Liquidity indicators calculated directly from transaction data often show bias, generally appearing better than the actual situation. We can reveal these behaviors by analyzing transaction relationships between addresses and identifying entity-linked addresses. Entity-linked address identification is closely related to meme token liquidity risk assessment. Through this technology, we can more accurately assess the true liquidity of tokens, excluding the influence of artificial manipulation factors such as self-trading (transactions between different addresses controlled by the same entity to artificially inflate trading volume without changing actual token ownership) and circular trading (sequential transactions among multiple entity-linked addresses forming a closed loop where tokens ultimately return to their original source, creating an illusion of distributed market activity while maintaining concentrated control). The contributions of this paper are as follows.
\begin{itemize}
\item Proposing an innovative multi-dimensional meme token related address identification method specifically for meme token market analysis. This method integrates four key dimensions: fund source analysis, fund destination analysis, behavior similarity analysis, and abnormal transaction behavior analysis, significantly improving the accuracy and comprehensiveness of entity-linked address identification.
\item Designing and implementing a comprehensive set of meme token liquidity risk assessment indicators, including three categories: token distribution, trading activity, and liquidity. This indicator system provides a theoretical framework for liquidity risk assessment in the meme token market.
\item Verifying the effectiveness of the liquidity risk assessment method based on entity-linked address identification through empirical analysis of meme tokens such as BabyBonk, NMT, and BonkFork. Experimental results show that our method can provide more accurate and reliable liquidity risk assessment for the meme token market based on real data, demonstrating significant application value in actual investment decisions.
\end{itemize}
Our work fills a significant gap in existing literature on meme token liquidity risk assessment, providing the first comprehensive analytical framework for this rapidly evolving and highly volatile specific cryptocurrency market category. The outcomes of this research are expected to significantly improve market transparency, reduce speculative bubbles, and contribute to building a healthier, more sustainable meme token ecosystem.

\section{Related Works}
Our research is the first comprehensive study specifically focused on assessing the liquidity risk of meme tokens. While significant progress has been made in blockchain address clustering, cryptocurrency liquidity analysis, and market manipulation detection, previous research has not specifically addressed liquidity risk. Our work builds upon and extends three main research areas, innovatively applying them to the meme token market.

\subsection{Entity Identification Blockchain Address Clustering and Entity Identification}

Blockchain address clustering plays a crucial role in understanding the true nature of cryptocurrency transactions and ownership. Victor \cite{b5} and Chen et al. \cite{b6} laid the foundation for address clustering and network analysis on Ethereum. In recent years, machine learning methods have been widely applied in this field, with works by Anouar et al. \cite{b7} and Wu et al. \cite{b8} significantly improving the accuracy of address clustering. Camino et al. \cite{b9}, Wang et al. \cite{b10}, and Zhong et al. \cite{b11} explored new methods to identify specific types of addresses and behavior patterns. Additionally, Ostapowicz and Żbikowski \cite{b12}, and Goldsmith et al. \cite{b13} made important advances in identifying suspicious transactions and illicit activities. Our work is the first to apply address clustering techniques to the meme token market, developing methods specifically tailored to the characteristics of meme tokens.

\subsection{Cryptocurrency Liquidity Analysis}
Cryptocurrency market liquidity analysis has seen significant developments. Research by Bidler et al. \cite{b14} and Kitzler et al. \cite{b15} provided new perspectives on understanding the liquidity dynamics of cryptocurrency markets. Zheng et al. \cite{b16} and Li and Yi \cite{b17} proposed new liquidity measurement methods, while Deng et al. \cite{b18} and Kondor et al. \cite{b19} focused on the liquidity characteristics of DeFi ecosystems and exchange networks. Ante \cite{b20}, Patel et al. \cite{b21}, and Kyriazis et al. \cite{b22} explored the liquidity formation process and value assessment methods for emerging tokens. 

Building on these works, our research is the first to propose a liquidity risk assessment framework specifically for meme tokens. We considered the unique features of meme tokens, such as high volatility and community-driven nature, to develop indicators and methods more suitable for assessing the liquidity risk of these tokens\cite{b3}.

\subsection{Market Manipulation Detection and Prevention}
Studies by Kamps and Kleinberg \cite{b23} and Cong et al. \cite{b24} provided important insights into identifying and understanding manipulative behaviors in cryptocurrency markets. Bian et al. \cite{b25} and Grasso et al. \cite{b26} proposed new methods for detecting abnormal trading patterns and wash trading. In terms of regulation, research by Chohan \cite{b27}, Zetzsche et al. \cite{b28}, and Cumming et al. \cite{b28} explored the regulatory challenges posed by meme tokens and DeFi markets. Furthermore, work by Ante and Meyer \cite{b30}, Shanaev et al. \cite{b31}, and Corbet et al. \cite{b32} provided new perspectives on understanding speculative behavior and bubbles in cryptocurrency markets. Liu et al. \cite{b33}, Zheng et al. \cite{b34}, and Huang et al. \cite{b35} made progress in cryptocurrency price prediction and risk assessment. Based on these studies, we propose using entity-linked addresses to reveal potential manipulative behaviors in meme tokens.

\begin{figure*}[htbp]
    \centering
    \includegraphics[width=0.8\textwidth]{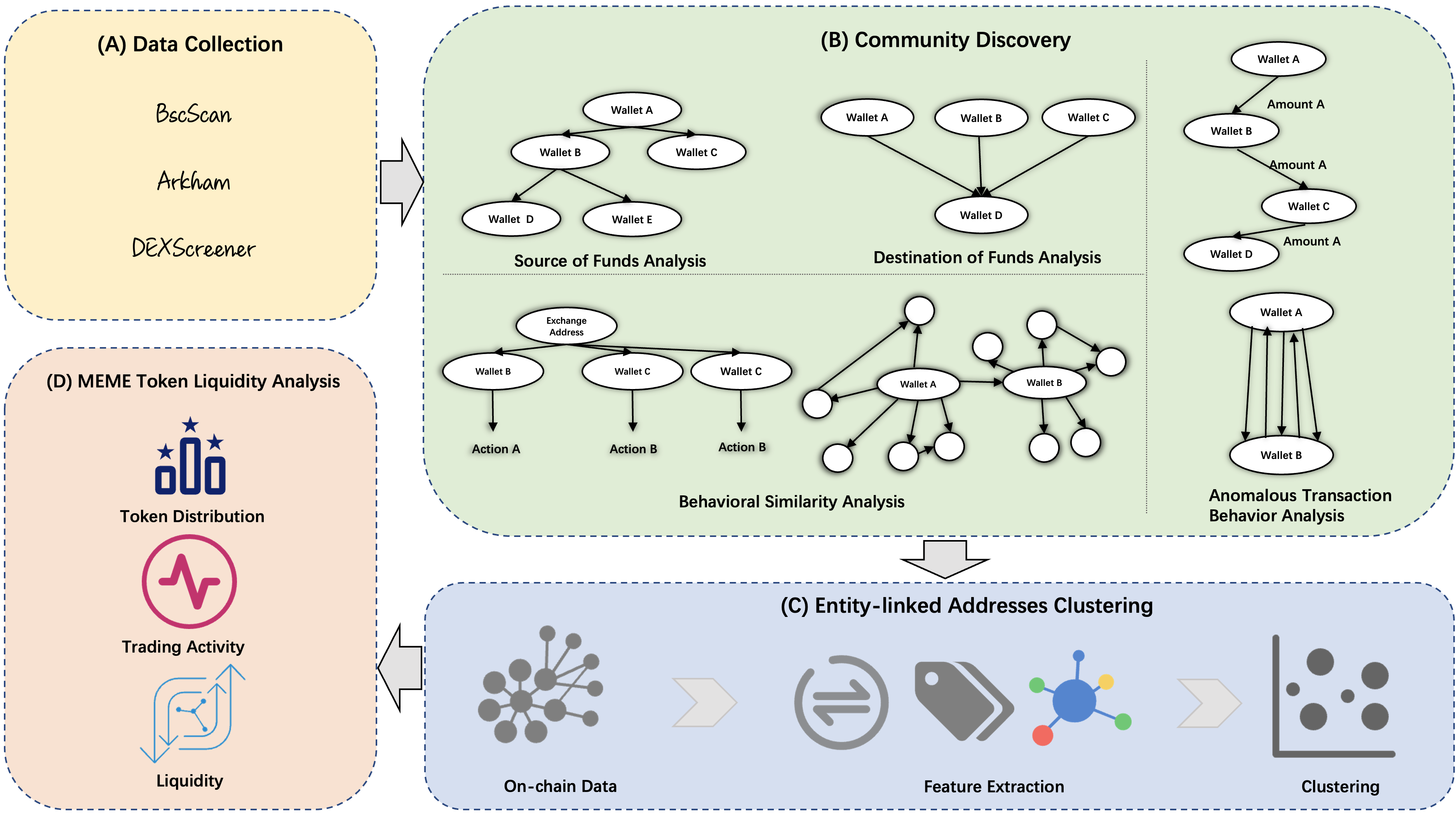}
    \caption{Workflow of the proposed liquidity risk analysis. It comprises the following steps: (A) Gathering data from BscScan, Akrham, and DEXScreener; (B) Utilizing Source of Funds Analysis, Destination of Funds Analysis, Behavioral Similarity Analysis, and Anomalous Transaction Behavior Analysis to identify entity-linked addresses; (C) Entity-linked Address Clustering and Merging: clustering, refining, and merging of the entity-linked addresses identified in (B); (D) Liquidity Indicator Calculation: Using the entity-linked address to calculate three categories of liquidity indicators: Token Distribution, Trading Activity, and Liquidity.}
    \label{fig:workflow_overview}
\end{figure*}


\section{Methodology}

As illustrated in Fig.~\ref{fig:workflow_overview}, we present the workflow. As shown in Fig.~\ref{fig:workflow_overview}(A), the on-chain data of the blockchain is sourced from BscScan(https://bscscan.com), the address label data is obtained from Arkham(https://intel.arkm.com), and the liquidity on decentralized exchanges is from DEXScreener(https://dexscreener.com). Building on this, we first introduce the use of graph mining algorithms to identify entity-linked addresses in Fig.~\ref{fig:workflow_overview}(B). Then, we describe how we filter outliers through additional clustering of features of addresses to improve precision in Fig.~\ref{fig:workflow_overview}(C). Finally, we introduce our proposed liquidity risk indicators in Fig.~\ref{fig:workflow_overview}(D). 

\subsection{Data Preprocessing} \label{sec:3.1}
The following two types of transactions
will be removed during the data preprocessing stage to avoid
incorrect entity linking: 
\begin{itemize}
\item Transactions involving publicly identifiable addresses, such as smart contract addresses and hot wallet addresses, are labeled using Arkham. It’s notable that fund transfers from hot wallets to multiple recipients don’t necessarily mean entity-linked ownership. For instance, since a hot wallet address is a shared address, if two addresses both receive funds from the same hot wallet, it doesn't imply any relationship between these two addresses. 

\item Transactions involving airdrop campaigns. Project-associated addresses (including token deployment and initial distribution) and multi-send contract addresses, whose labels are sourced from Arkham, are used to identify airdrop campaigns. The identification rules for airdrop campaigns transactions are as follows: (1) Transfers of similar amounts of funds are made from project-associated addresses to multiple addresses within a single transaction. (2) Airdrop campaigns often use multi-send contracts to transfer funds from one address to many addresses within a single transaction.
\end{itemize}


\subsection{Community Discovery} \label{sec:3.2}
As depicted in Fig.~\ref{fig:workflow_overview}(B), we employ four distinct methods to identify entity-linked addresses. These methods complement each other, with each method targeting different types of blockchain behaviors and transaction patterns. By comprehensively applying these methods and performing a final aggregated analysis of the results, we can obtain a more comprehensive and accurate identification of entity-linked addresses. 
\subsubsection{Source of Funds Analysis} 
This method is based on a reasonable assumption: funds originating from the same address (Exclude institutional addresses such as smart contract addresses and hot wallet addresses) are likely to belong to the same entity. We primarily focus on two typical patterns: the diffusion funding pattern and the sequential diffusion funding pattern. The diffusion funding pattern refers to a main wallet transferring funds to multiple sub-wallets, forming a "one-to-many" fund flow graph. The sequential diffusion funding pattern is characterized by funds being transferred sequentially between a series of wallets, forming a chain-like fund flow path. The main advantage of this method lies in its high reliability. For address clusters with the same funding source, we can be highly confident that they belong to the same entity. This method is particularly effective for holding addresses with low-frequency transactions and fewer counterparties. Importantly, this method does not need to consider time similarity, making it especially useful for analyzing the behavior of long-term holders. 
\subsubsection{Destination of Funds Analysis}
Multiple addresses belonging to the same person transfer funds to a single address. This "many-to-one" fund flow typically indicates that these small-value addresses may belong to the same entity. The advantage of this method is that it can overcome issues related to addresses with limited operational behavior or lack of time similarity. In the meme token market, this method can help us detect market manipulation by large holders using multiple small addresses, thereby more accurately assessing liquidity risk. This method is particularly useful in identifying behaviors that attempt to conceal the true scale of assets through dispersed holdings. For example, a large holder might use multiple small-value addresses to hold and trade tokens to avoid drawing attention.
\subsubsection{Behavioral Similarity Analysis}
This method is used to address complex situations where entity-linked addresses attempt to evade traditional fund source detection algorithms. In Fig.~\ref{fig:workflow_overview}(B), the behavioral similarity analysis section features two subgraphs representing two common patterns. The left subgraph shows that some entity-linked addresses may directly fund from centralized exchanges to avoid direct fund associations. However, their subsequent operation patterns often show similarities, such as having similar transaction times or interacting with the same contract. Meanwhile, the right subgraph illustrates that these addresses belong to the same entity-linked addresses due to their interconnected transaction relationships. We employ the Louvain community detection algorithm\cite{b36} to effectively identify entity-linked address groups with similar operation patterns and transaction timing. The advantage of this method lies in its ability to capture more complex patterns of entity-linked addresses. In the meme token market, this analysis can help us discover groups operating in coordination, which may be attempting to influence market prices or liquidity through dispersed operations.
\subsubsection{Anomalous Transaction Behavior Analysis}
This method primarily focuses on two suspicious transaction patterns: transfers of almost identical amounts between different addresses, and high-frequency transactions among multiple addresses. As a complementary method, it can capture special cases that might be overlooked by other methods, thus providing an additional layer of verification. An important application of this method is identifying "circular trading" or "self-trading" behaviors. These behaviors may aim to artificially increase transaction volume or manipulate prices. By detecting round-trip transactions with identical amounts or high-frequency transactions with small amounts, we can reveal these potential market manipulation behaviors. This is crucial for assessing the true liquidity and market activity of meme tokens, as the markets for these tokens are typically more susceptible to such behaviors.

\subsection{Entity-linked Addresses Clustering} \label{sec:3.3}
\subsubsection{Feature Extraction and Clustering}
In the process of identifying entity-linked addresses, cluster analysis after community detection is a crucial step. Community detection provides the macrostructure of entity groups, while clustering filters out outlier addresses within the entity groups identified by community detection by considering features such as transaction patterns, time series, and contract interactions. Feature extraction is key to the clustering process, and we extract features from multiple dimensions, mainly including: 
\begin{itemize}
    \item Basic Transaction Features: Transaction frequency, amount distribution, number of transactions, gas\_fee, etc. 
    \item Network Topology Features: The position and importance of addresses in the transaction network. 
    \item Time Series Features: Patterns of transaction behavior over time. 
    \item Token Holding Features: Types, quantities and durations of held tokens. 
    \item Social Graph Features: Characteristics based on transaction counterparts. 
    \item Historical Label Information: Known address type labels.

\end{itemize}
Our clustering methodology employs DBSCAN\cite{b37} to identify preliminary transaction-based communities, followed by isolation forest algorithm\cite{b38} to remove anomalous addresses while maintaining community integrity. The refined groups undergo evaluation through a probabilistic model that assesses transaction patterns, address similarities, token flows, and temporal correlations, with the final entity-linked address groups determined by applying preset probability thresholds.

\subsection{MEME Token Liquidity Metrics} \label{sec:3.4}
After identifying entity-linked addresses, we can calculate liquidity indicators for meme tokens through these addresses, which better reflect their true liquidity conditions. We selected six indicators across three categories to describe token liquidity, based on several considerations, including the existing data platforms for meme tokens, discussions with industry professionals from leading exchanges, and the need to reflect both static and dynamic aspects of liquidity:

\begin{itemize}
    \item Optimized Token Distribution Indicators
    \begin{itemize}
        \item Top 10 Position: Percentage of tokens held by the top 10 holders. This indicator helps to understand the concentration of token ownership, which is crucial for assessing market stability and potential risks. The distribution is calculated at the current moment to reflect the most recent ownership structure.
        \item Herfindahl-Hirschman Index (HHI)\cite{b39}: An economic indicator used to measure market concentration. HHI provides a quantitative measure of the distribution of token holdings, helping to identify potential monopolistic or oligopolistic market structures. The HHI is also calculated at the current moment.
        $$
          HHI = \sum_{i=1}^{n} p_i^2
        $$
        where $p_i$ is the market share of the $i$-th holder.
    \end{itemize}
    \item Optimized Trading Activity Indicators
    \begin{itemize}
        \item VMTV (24-hour trading volume to market cap ratio)\cite{b40}. This ratio helps to assess the liquidity and market activity of the token, providing insights into how liquid the token is relative to its market capitalization. The trading volume and market cap are calculated over the most recent 24-hour period.
        $$
          VMTV = \frac{V}{MC}
        $$
        where $V$ is the 24-hour trading volume and $MC$ is the market capitalization.
        \item Volatility (24-hour trading volume to liquidity ratio): This indicator reflects the token’s turnover rate, indicating how frequently tokens are traded and helping to identify potential manipulation or artificial trading activities. Both the trading volume and liquidity are calculated over the most recent 24-hour period.
        $$
            \text{Volatility} = \frac{V}{L}
        $$
        Here, $V$ is the 24-hour trading volume, representing the total amount of tokens traded within the last 24 hours. $L$ is the total value of tokens available in the liquidity pool at the current moment.
    \end{itemize}
    \item Optimized Liquidity Indicators
    \begin{itemize}
        \item Liquidity: The liquidity on Decentralized Exchanges (DEX) is measured by the total value of all tokens in the liquidity pool, reflecting the amount of capital available for trading. This metric provides a quantitative measure of the liquidity available in the pool, which is crucial for assessing the token's tradability and market depth. Assuming the liquidity pool consists of two tokens, Token A and Token B, with quantities $Q_A$ and $Q_B$, and current prices $P_A$ and $P_B$, respectively. The total value $L$ of the liquidity pool can be expressed as:
        $$
            L = Q_A \cdot P_A + Q_B \cdot P_B
        $$
        \item Holders: The number of token holders provides insights into the token’s community engagement and potential for future growth. This metric is calculated by counting the unique addresses holding the token at the current moment.
        
    \end{itemize}
\end{itemize}
The liquidity indicators, optimized through the identification of entity-linked addresses, more effectively reflect the true market conditions and potential risks of meme tokens. The optimized Top 10 holdings and HHI indicators more accurately reveal the actual concentration of funds and market competitiveness. The optimized VMTV and volatility indicators exclude artificially generated false trading activities, providing a more authentic assessment of market liquidity. Simultaneously, the optimized holder indicators present a more accurate distribution of market participants.

\section{Experiment}
\subsection{Data Collection and Preparation}
Meme tokens experienced significant price increases in March, and the positive performance of these token prices has sparked our interest in the true value of these memes. As shown in Fig.~\ref{fig:workflow_overview}(A), we demonstrate the effectiveness of our proposed method by analyzing the meme token BabyBonk (contract address 0xbb2826ab03b6321e170f0558804f2b6488c98775) , a high-ranking meme token on the BSC chain. We extracted transaction data for this token spanning from December 15, 2023, to March 23, 2024. This dataset encompasses 275,956 transactions and 21,759 holding addresses. We applied the data preprocessing methods outlined in Section~\ref{sec:3.1} to clean the transaction data: (1) Identification and removal transaction of publicly identifiable addresses: Using the Arkham API, we identified 43 smart contract addresses and 10 hot wallet addresses, which involved 130,741 transactions; (2) Exclusion of airdrop campaign transactions: We further identified 798 addresses that participated in airdrop campaigns, which involved 3,879 transactions. After excluding the aforementioned transactions, the final dataset consisted of 161,336 transactions and 18,587 holding addresses.

\subsection{Community Address Recognition}
After identifying entity-linked groups using the four methods mentioned in Section~\ref{sec:3.2}, we then apply the method described in Section~\ref{sec:3.3} to further refine and consolidate the addresses within these entity-linked groups. Then, we use entity-linked groups to calculate the liquidity indicators mentioned in Section~\ref{sec:3.4}. Finally, we compare the liquidity indicators before and after applying the entity-linked identification process, thereby demonstrating the effectiveness of our proposed method.
\subsubsection{Source of Funds Analysis}
By setting thresholds of minimum 5 receiving addresses and 10 USDT as the minimum transaction amount, we filter out noise, resulting in the discovery of 1,063 entity-linked groups encompassing 5,413 addresses. As shown in the figure below, the funds for these addresses come solely from the same address.



\subsubsection{Destination of Funds Analysis}
By setting thresholds of minimum 5 sending addresses and 10 USDT as the minimum transaction amount, we filter out noise, resulting in the discovery of 2,811 entity-linked groups encompassing 3,008 addresses.



\subsubsection{Behavioral Similarity Analysis}
By setting thresholds of 10 USDT as the minimum transaction amount, we filter out noise, resulting in the discovery of 1,104 entity-linked groups encompassing 5,819 addresses. As shown in the figure below, we have selected one of these entity-linked groups as an example. It can be observed that their transactions form a tree-like structure, where a central address distributes funds to a group of addresses, and these addresses then continue to distribute funds to other addresses.
\begin{figure}[htbp]
    \centering
    \includegraphics[width=\columnwidth]{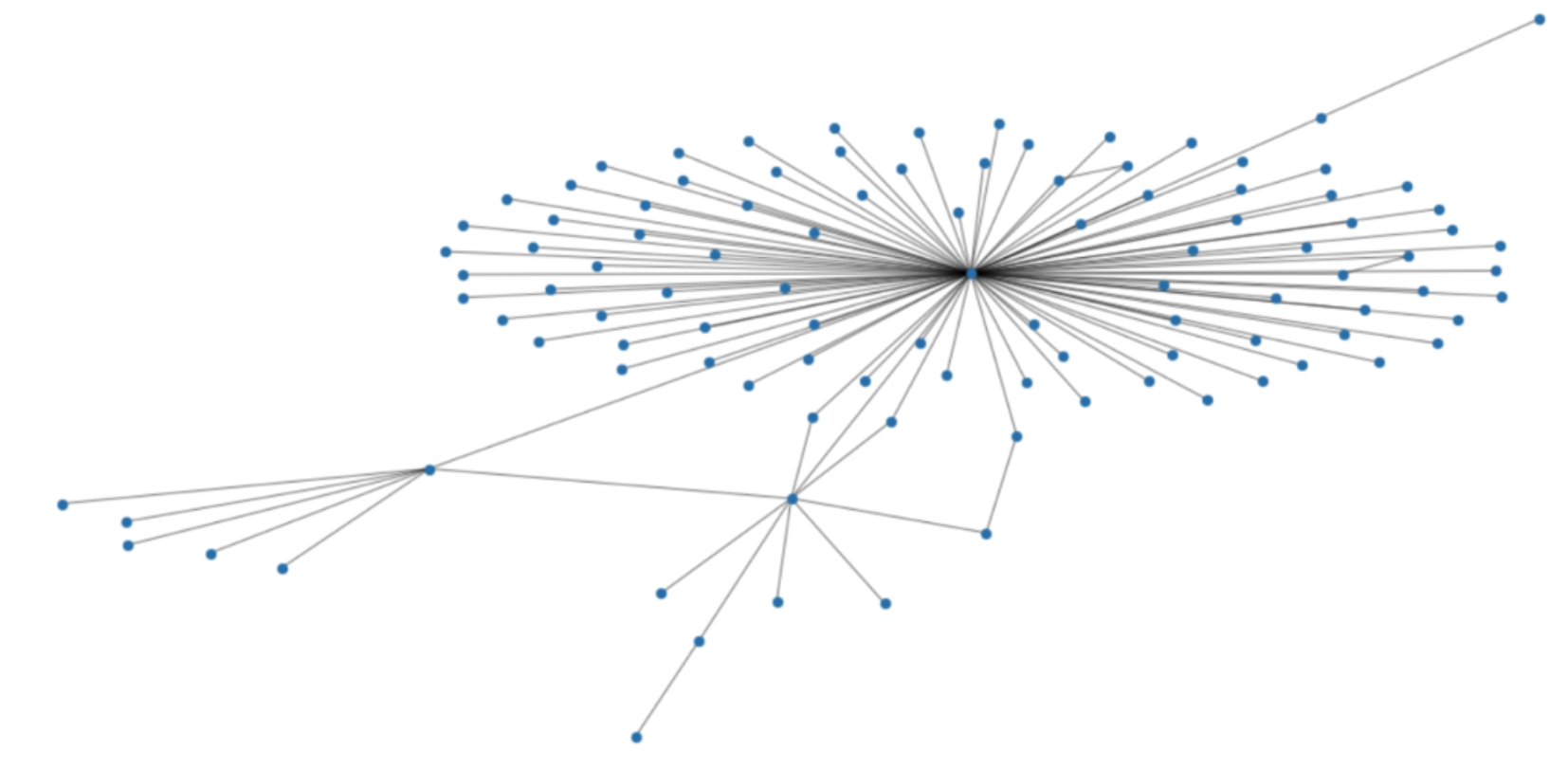}
    \caption{Tree-like fund distribution.}
    \label{fig:fig4}
\end{figure}

\subsubsection{Anomalous Transaction Behavior Analysis}
By setting thresholds of minimum 5 transactions and 5 USDT as the minimum transaction amount, we filter out noise, resulting in the discovery of 70 entity-linked groups encompassing 2,015 addresses. As shown in the figure below, we selected all transactions with a transfer amount of one hundred billion and plotted a transaction graph for the addresses involved in these transfers. From the transaction relationships, we can observe a strong correlation among these addresses.
\begin{figure}[htbp]
    \centering
    \includegraphics[width=\columnwidth]{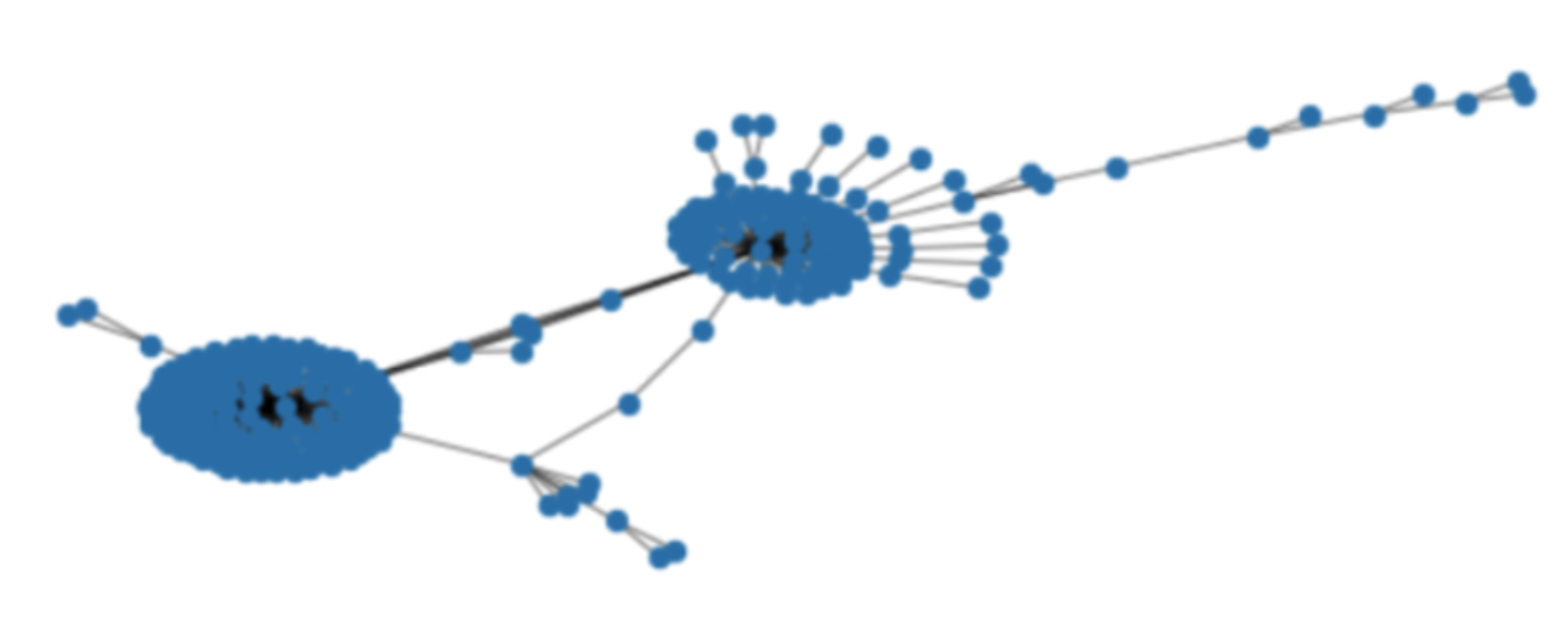}
    \caption{Transaction graph of identical fund amounts.}
    \label{fig:fig5}
\end{figure}

\subsubsection{Entity-linked Addresses Clustering Results}
\begin{figure}[htbp]
    \centering
    \includegraphics[width=\columnwidth]{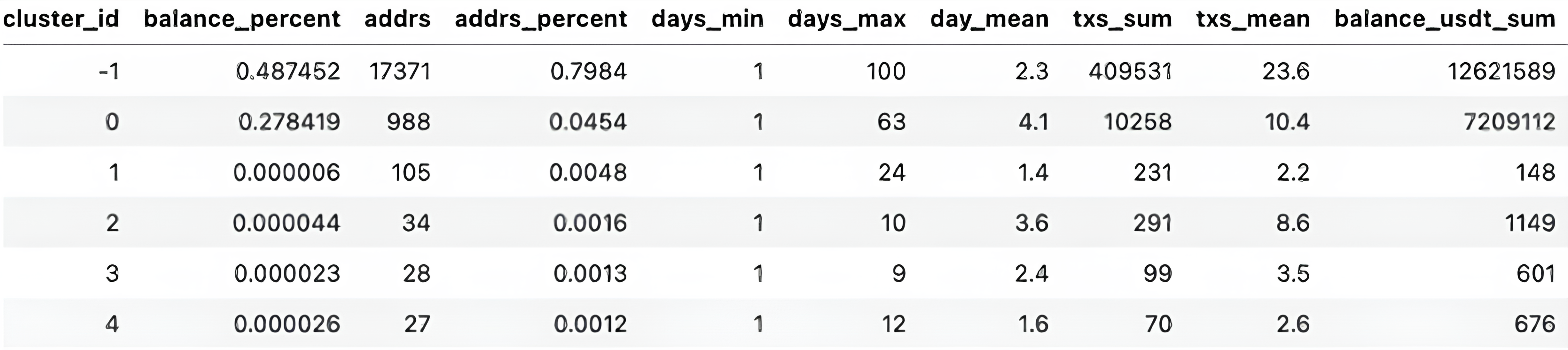}
    \caption{Statistics of the top 5 entity-linked groups by number of addresses.}
    \label{fig:fig6}
\end{figure}

\begin{figure}[htbp]
    \centering
    \includegraphics[width=\columnwidth]{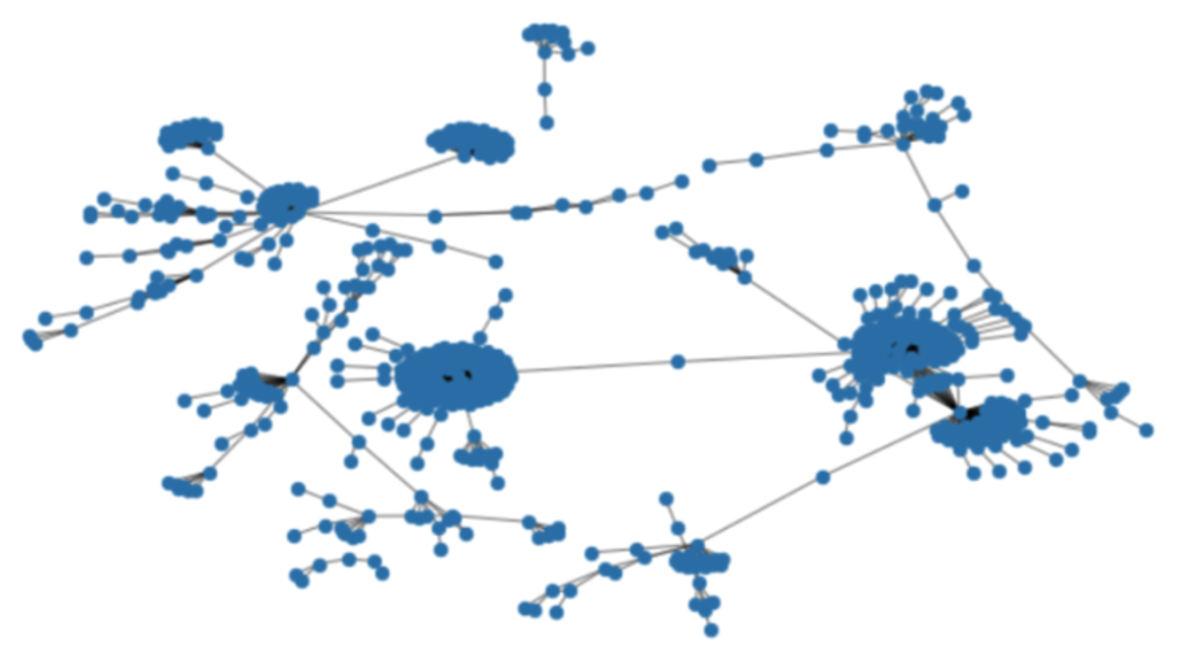}
    \caption{Entity-linked groups of cluster\_id = 0.}
    \label{fig:fig7}
\end{figure}

Employing a multistage analytical approach, we aggregated 18,587 distinct addresses into 5,245 entity-linked groups. This process comprised DBSCAN clustering (Eps=0.5, MinPts=5, chosen to balance cluster density and noise tolerance), Isolation Forest anomaly detection (contamination rate 0.1, based on empirical observations of outlier prevalence in data), and a probabilistic entity linkage model (threshold 0.7, set conservatively to minimize false positives). These parameters were optimized through iterative testing to maximize the accuracy of the entity identification. Finally, 1,214 entity-linked groups were identified, comprising a total of 4,387 addresses. The top 5 entity-linked groups in Fig.~\ref{fig:fig6} ranked by number of addresses are as follows:
\begin{itemize}
    \item The entity-linked holding 49\% of the total tokens has only one address (cluster label=-1), indicating many retail investors. A higher number of retail investors holding the token indicates good interest and attention towards the token.
    \item The entity-linked with the largest number of addresses (cluster\_id=0) holds 27.8\% of the tokens. Considering its 10,258 transactions and continuous trading activity since the token's inception, as well as the tightly knit transaction relationships observed in Fig.~\ref{fig:fig7}, this suggests it may be a market-making group. Therefore, it can be excluded from subsequent analyses.   
    
\end{itemize}

\begin{figure}[htbp]
    \centering
    \includegraphics[width=0.8\columnwidth]{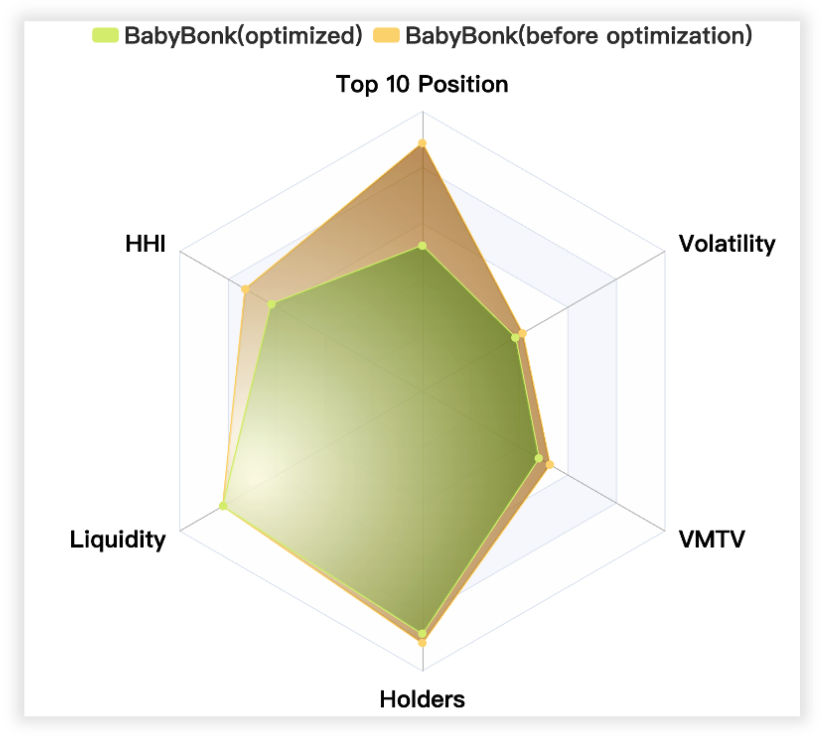}
    \caption{Comparison of BabyBonk token liquidity before and after entity-linked optimization.}
    \label{fig:fig8}
\end{figure}

\begin{figure*}[htbp]
    \centering
    \includegraphics[width=0.9\textwidth]{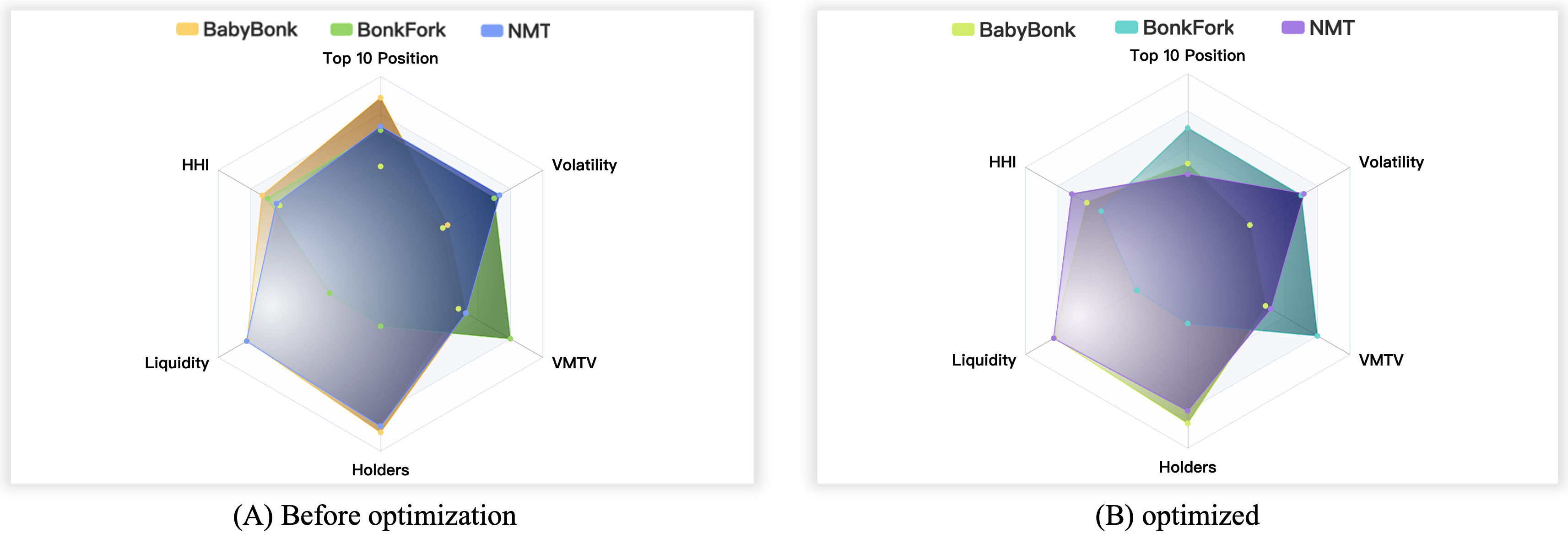}
    \caption{Comparison of BabyBonk/NMT/BonkFork token liquidity before and after entity-linked optimization.}
    \label{fig:fig9}
\end{figure*}

\subsection{MEME Token Liquidity Analysis}
We use the identified entity-linked groups to calculate the indicators mentioned in Section 3.3 and compare the results before and after optimization using entity-linked identification. We convert all these liquidity indicators into positive indicators, which means that higher values indicate better liquidity. As shown in Fig.~\ref{fig:fig8}, a larger area in the radar chart represents better liquidity. We can observe that after optimization using entity-linked groups for BabyBonk, all indicators except the liquidity indicator have decreased. The change in the Top 10 Position is the most significant. After entity-linked identification, we can discover that the actual concentration of the top 10 token holders is higher, revealing the true liquidity risk.

We selected three meme tokens on the Binance Smart Chain (BSC) for the period of March 2024: BabyBonk and NMT, which were among the top five ranked tokens on DexScreener during this period, and BonkFork, which exhibited high trading activity. We conducted a comparative analysis of their liquidity profiles to decide which token to purchase. BonkFork token's liquidity indicators were relatively poor, so it was eliminated first. As shown in Fig.~\ref{fig:fig9}(A), comparing BabyBonk and NMT before entity-linked optimization, NMT had better volatility indicators but worse top 10 position indicators, indicating higher trading activity but also higher token concentration. BabyBonk token exhibited opposite characteristics, making it difficult to decide which token to purchase. As shown in Fig.~\ref{fig:fig9}(B), after applying entity-linked optimization, we can see that the radar chart area for the NMT token is significantly larger than that for BabyBonk token, indicating better liquidity for NMT token. We included this token in our consideration set, and from our analysis of subsequent price trends, we observed that NMT demonstrated relatively higher price stability compared to other tokens.

\section{Limitations and Discussion}
Despite the significant progress made in this study on meme token liquidity risk analysis, several limitations remain. Firstly, the accuracy of entity-linked address identification may be affected by complex strategies, and excluding public addresses might lead to overlooking some genuine associations. Secondly, the current static indicator system may not fully capture the dynamic changes in the market, especially considering the highly speculative nature and rapid fluctuations characteristic of the meme token market.

Additionally, our study lacks longitudinal validation of how identified risk factors correlate with long-term price movements. While our analysis indicated that tokens like NMT exhibited higher price stability, a more comprehensive examination of price trends over extended periods would strengthen our validation framework. Meme token price movements are influenced by a complex array of factors, of which liquidity considerations represent only one dimension. Our analytical framework deliberately focuses on liquidity risk metrics—particularly in identifying potential rug-pull vulnerabilities where seemingly diversified token distributions may actually conceal concentrated holdings among interconnected insider addresses—but social media engagement, market sentiment, and ideological alignment with the token's narrative all exert significant influence on price trajectories beyond our current methodology.

Furthermore, potential selection bias exists in our empirical analysis, as the tokens chosen may not fully represent the broader meme token market, potentially limiting the generalizability of our findings. In the data cleaning process, the accuracy of identifying and excluding project team airdrop activities also poses a challenge for comprehensive risk assessment.

\section{Conclusion and Future Work}

Our research introduces a novel framework for assessing liquidity risk in meme token markets through entity-linked address identification techniques. The main achievements of our study are: 1) The effectiveness of our multidimensional entity-linked address identification method in revealing hidden relationships within meme token transactions, providing a more accurate picture of market dynamics; 2) The revelation of significant discrepancies between apparent and actual liquidity in meme token markets, emphasizing the importance of our approach in risk assessment; 3) The successful application of our liquidity risk assessment indicators to meme tokens, demonstrating their practical utility in investment decision-making. These findings provide valuable insights for investors, market analysts, and regulators, contributing to a more transparent and stable meme token ecosystem.

Future work will focus on enhancing both risk assessment capabilities and investment guidance through incorporating longer-term price trajectory analysis. This empirical validation will help determine whether our risk indicators effectively predict market outcomes, providing more robust evidence regarding their practical utility. Our goal is to develop a dynamic indicator system that can better correlate with price trends while providing clear and interpretable conclusions to users. This will involve designing a weighted scoring system for liquidity risk, implementing adaptive thresholds, and developing time-based indicators to track liquidity trends over extended periods.

During the writing of this paper, we chose meme tokens on Binance Smart Chain (BSC) mainly due to their relatively high popularity during that period. Our proposed entity recognition method is applicable to other active public chains of meme tokens such as Ethereum (ETH), BASE, and Solana, although the data pre-processing methods vary for different chains. The core concept of the entity-linked address recognition method remains valid across these platforms, providing a foundation for analyzing meme tokens in the broader ecosystem.




\end{document}